# An enhanced version of the Gaia map of the brightness of the natural sky

Eduard Masana[1,*], Salvador Bará[2],

Josep Manel Carrasco[1] and Salvador J. Ribas[3,1]

[1]Departament Física Quàntica i Astrofísica. Institut de Ciències del Cosmos (ICC-UB-IEEC), C Martí Franquès 1, Barcelona 08028, Spain
[2]Departmento Física Aplicada, Universidade de Santiago de Compostela, 15782 Santiago de Compostela, Galicia, Spain
[3]Parc Astronòmic Montsec - Ferrocarrils de la Generalitat de Catalunya. Camí del Coll d'Ares, s/n, Àger (Lleida) 25691, Spain

**Abstract**

The GAia Map of the Brightness Of the Natural Sky (GAMBONS) is a model to map the natural night brightness of the sky in cloudless and moonless nights. It computes the star radiance from the photometric data in Gaia and Hipparcos catalogues, adding the contributions of the diffuse galactic and extragalactic light, zodiacal light and airglow, and taking into account the effects of atmospheric attenuation and scattering. The model allows computing the natural sky brightness in any given photometric band for a ground-based observer, if appropriate transformations from the Gaia bands are available. In this work we present the most recent improvements of the model, including the use of Gaia EDR3 data, the inclusion of a wide set of photometric bands and the derivation of additional sky brightness indicators, as the horizontal irradiance and the average hemispheric radiance.

## 1. Introduction

Models of the sky brightness, both its natural and its artificial components, are powerful tools helping to establish a baseline against which to evaluate the light pollution levels, interpret the measurements and predict future scenarios of artificial lighting.

The GAia Map of the Brightness Of the Natural Sky (GAMBONS, [1], referred as Paper I hereinafter) was published with the aim to be used as a reference to assess the quality of the night sky at any given place. The possibility of setting the atmospheric conditions (in particular the aerosol optical depth, directly related to the atmospheric attenuation) and the airglow intensity, as well as its multi-band capability, allow to compute accurate all-sky maps in different photometric passbands. The first version of the model included the Johnson V band, the Gaia G band and the photopic and scotopic human vision bands. On the other hand, besides the detailed full-resolution sky maps, the only synoptic sky brightness indicator provided by the model was the zenith magnitude (defined as the magnitude, in mag arcsec$^{-2}$, corresponding to the average radiance within a circular region of the sky around the zenith). We present in this work an updated version of GAMBONS, which includes new photometric passbands and additional sky brightness indicators. This new version also recomputes the integrated starlight using the Early Data Release 3 of the Gaia mission (Gaia EDR3), a new catalogue with a substantial improvement of the photometric data with regards to the Data Release 2 (DR2) used in the previous version of our model.

---

* E. Masana, *E-mail address:* emasana@fqa.ub.edu

For a summary of previous works on modelling the natural sky brightness and a detailed description of GAMBONS, we refer the reader to Paper I.

## 2. Gaia EDR3 data

The first version of GAMBONS was based on Gaia DR2 data. After its publication, a new data Gaia catalogue (Gaia EDR3) has been released. Gaia EDR3 catalogue contains astrometry and photometry for 1.8 billion sources brighter than magnitude G=21 mag. For 1.5 billion sources, the (GBP−GRP) color is also available. The Gaia EDR3 photometry is a substantial improvement from Gaia DR2, specially at the bright end (G < 13), where many of the systematics effects reported in Gaia DR2 have been removed or greatly suppressed. Furthermore, a single passband for each G, GBP, and GRP is band valid over the entire magnitude and colour range, with no systematics above the 1 percent level [2-3]. The typical uncertainty of the Gaia EDR3 photometry at G=17 mag. is σG = 0.8 mmag for G, σGBP = 5 mmag for GBP and σGRP =7.5 mmag for GRP band.

As in Paper I, for the bright stars not included in Gaia we have used Hipparcos data. The computation of the integrated starlight from Gaia and Hipparcos data follows the same methodology described in Paper I. The photometric transformations have been recomputed to account for the new definition of the Gaia EDR3 filters [3].

A new catalogue of the radiance outside the Earth atmosphere is available online for all the bands described in the next section.

Table 1. Main characteristics of GAMBONS implemented photometric bands. See text for the column's description.

| System id. | Passband | $\lambda_m$ | $\mathcal{A}$ | $L_r$ | $\langle L_{r,\lambda} \rangle$ | Radiance units | Magnitude scale |
|---|---|---|---|---|---|---|---|
| | | [nm] | | [Rad. Units] | [Rad. units· nm$^{-1}$] | | |
| Johnson – Cousins | $U$ | 361.6 | 60.5 | 107.340 | 1.773 | W m$^{-2}$ sr$^{-1}$ | Vega |
| Johnson – Cousins | $B$ | 441.9 | 95.9 | 264.689 | 2.759 | W m$^{-2}$ sr$^{-1}$ | Vega |
| Johnson – Cousins | $V$ | 552.4 | 90.9 | 143.168 | 1.575 | W m$^{-2}$ sr$^{-1}$ | Vega |
| Johnson - Cousins | $R$ | 660.8 | 163.4 | 153.817 | 0.942 | W m$^{-2}$ sr$^{-1}$ | Vega |
| Johnson - Cousins | $I$ | 802.0 | 147.5 | 74.115 | 0.503 | W m$^{-2}$ sr$^{-1}$ | Vega |
| *Gaia* | $G$ | 639.0 | 317.3 | 1.119 10$^{21}$ | 3.527 10$^{18}$ | photons s$^{-1}$m$^{-2}$ sr$^{-1}$ | Vega |
| SDSS | $u$ | 356.2 | 6.1 | 3.978 10$^{19}$ | 6.565 10$^{18}$ | photons s$^{-1}$m$^{-2}$ sr$^{-1}$ | AB |
| SDSS | $g$ | 471.9 | 41.9 | 2.082 10$^{20}$ | 4.975 10$^{18}$ | photons s$^{-1}$m$^{-2}$ sr$^{-1}$ | AB |
| SDSS | $r$ | 618.5 | 54.6 | 2.065 10$^{20}$ | 3.781 10$^{18}$ | photons s$^{-1}$m$^{-2}$ sr$^{-1}$ | AB |
| SDSS | $i$ | 750.0 | 44.2 | 1.378 10$^{20}$ | 3.117 10$^{18}$ | photons s$^{-1}$m$^{-2}$ sr$^{-1}$ | AB |
| SDSS | $z$ | 896.1 | 8.9 | 2.310 10$^{19}$ | 2.610 10$^{18}$ | photons s$^{-1}$m$^{-2}$ sr$^{-1}$ | AB |
| RGB | $B$ (Blue) | 469.1 | 84.7 | 4.230 10$^{20}$ | 4.994 10$^{18}$ | photons s$^{-1}$m$^{-2}$ sr$^{-1}$ | AB |
| RGB | $G$ (Green) | 532.4 | 91.7 | 4.037 10$^{20}$ | 4.404 10$^{18}$ | photons s$^{-1}$m$^{-2}$ sr$^{-1}$ | AB |
| RGB | $R$ (Red) | 600.7 | 68.5 | 2.674 10$^{20}$ | 3.904 10$^{18}$ | photons s$^{-1}$m$^{-2}$ sr$^{-1}$ | AB |
| SQM | SQM | 516.0 | 222.8 | 1.037 10$^{21}$ | 4.654 10$^{18}$ | photons s$^{-1}$m$^{-2}$ sr$^{-1}$ | AB |
| TESS-W | TESS-W | 566.5 | 326.1 | 1.391 10$^{21}$ | 4.251 10$^{18}$ | photons s$^{-1}$m$^{-2}$ sr$^{-1}$ | AB |
| Human vision | $V'$ Scotopic | 502.4 | 97.1 | 181.204 | 1.867 | W m$^{-2}$ sr$^{-1}$ | AB |
| Human vision | $V$ Photopic | 560.2 | 106.9 | 160.417 | 1.501 | W m$^{-2}$ sr$^{-1}$ | AB |

## 3. Photometric Bands

In the new GAMBONS version described here, we have implemented several new photometric bands. For all of them we have computed the transformations from Gaia ($G$ and GBP-GRP) and Hipparcos ($V$ and $B - V$ or $V - I$) photometry to the given band. Note that due to the spectral coverage of the Gaia and Hipparcos photometry, the photometric transformations are less accurate for the bluer bands, as Johnson $U$ or Sloan $u$ bands. Figure 1 shows the transmission curves and Table 1 summarizes their main characteristic:

- Mean wavelength $\lambda_m$, defined as the weighted average:

$$\lambda_m = \frac{\int_0^\infty \lambda\ S(\lambda)\ d\lambda}{\int_0^\infty S(\lambda)\ d\lambda} \qquad (1)$$

- $\mathcal{A} = \int_0^\infty S(\lambda)\ d\lambda$, the area below the passband curve $S(\lambda)$ expressed in nm.

- Reference radiance $L_r$ used to transform from the in-band radiance of the sky, $L$, to magnitudes per arcsec$^2$ as $m = -2.5\log(L/L_r)$. $L_r$ is calculated as the band-weighted integral of the reference spectral radiance $L_{r,\lambda}$ chosen to define the magnitude scale (see below):

$$L_r = \int_0^\infty L_{r,\lambda}\ S(\lambda)\ d\lambda \qquad (2)$$

The units of $L$ and $L_r$ are photons s$^{-1}$m$^{-2}$sr$^{-1}$ for photon-based radiometric quantities, or W m$^{-2}$ sr$^{-1}$ for energy-based ones.

- Band-averaged reference spectral radiance, $\langle L_{r,\lambda}\rangle = L_r/\mathcal{A}$, expressed in photons s$^{-1}$m$^{-2}$sr$^{-1}$nm$^{-1}$ or W m$^{-2}$sr$^{-1}$nm$^{-1}$.

- Magnitude scale adopted for the definition of the radiometric zero points. The Vega scale is defined consistently with the usual choice of +0.03 mag for the stellar magnitude of Vega in all photometric bands. In the Vega scale $L_r$ is then the in-band radiance of a source of angular extent 1 arcsec$^2$ that would produce $10^{0.4 \times 0.03}$ times the STIS003 spectral irradiance of Vega [4], at the entrance pupil of the observing instrument under normal incidence. The AB scale is defined in a similar way, being in this case $L_r$ the in-band radiance of a source of 1 arcsec$^2$ angular extent that would produce the AB spectral irradiance $F_\nu = 3.63\ 10^{-23}$ W m$^{-2}$ Hz$^{-1}$ [5] over the entrance pupil of the instrument.

*3.1. Johnson – Cousins bands*

The UVBRI Johnson-Cousins system [6-8] has been the more widely used photometric system in astronomy along the years. It covers wavelengths from $\approx$ 300 nm to $\approx$ 900 nm.

In light pollution studies, the night sky brightness at the $V$ band has been traditionally used to describe the sky brightness quality [9], even when virtually no devices (with the exception of ASTMON cameras [10]) are natively equipped with this filter. The Johnson-Cousins $V$ band is also commonly (and incorrectly) identified with the spectral sensitivity of the human vision. On the other hand, the characterization of the sky brightness in the $B$, $V$ and $R$ bands, and their associated colors, has been used to study the effect of the different lighting systems (LEDs, high pressure sodium lamps, mercury vapor lamps,...) on the sky brightness [11].

*3.2. Gaia band*

Gaia EDR3 catalogue is the largest and most accurate catalogue of photometric and astrometric data published to the date, with more than 1.8 billion sources [2]. The Gaia broad band photometric system is fully described in [12]. It is composed by three different passbands ($G$, GBP and GRP). The wide $G$ passbands covers a wavelength range from $\approx$ 330 nm to $\approx$ 1050 nm. Although Gaia photometric system is not yet used for ground based observations, it is the current needed reference for astrophysical sources and we have included the $G$ Gaia in GAMBONS.

*3.3. Sloan Digital Sky Survey bands*

Sloan Digital Sky Survey (SDSS) photometric system [13] is composed by five mostly non-overlapping filters ($u$, $g$, $r$, $i$ and $z$), with wavelength coverage from 300 nm (atmospheric ultraviolet cutoff) to 1100 nm (the sensitivity limit of silicon CCDs). SDSS has been observing the skies from Apache Point Observatory since 1998 and from Las Campanas Observatory since 2017. The twelfth data release [14] of the SDSS archive contains more than 469 million sources, 260 millions of them classified as stars. The Panoramic Survey Telescope and Rapid Response System (Pan-STARRS, [15]) and the Large Synoptic Survey Telescope (LSST, [16]) photometric systems, among other large photometric surveys, have optical bands very similar, but not identical, to the SDSS bands. Although the use of the SDSS photometric system is not yet common in light pollution studies, this situation could change in the future, as its use is increasing in all astronomy fields in the last years.

*3.4. The SQM and TESS-W bands*

The Sky Quality Meter (SQM) of Unihedron (Grimsby, ON, Canada) [17-18] is a night sky brightness meter based on an irradiance-to-frequency converter chip. It is the more widely used instrument to take field measurements in light pollution research, as well as to monitor the evolution of the sky brightness over the time in a given place. The SQM passband is determined by the optics transmission and the response of its chip. As it is shown in Fig. 1 ([19], [18]), it is a broad band covering from about 400 nm to 1000 nm.

The TESS-W photometer is the first version of the Telescope Encoder and Sky Sensor sky brightness meter [20,18]. This device also measures the ambient and sky temperature to record the cloud coverage. Its sensitivity is higher than that of the SQM, and its spectral response is extended to the red, up to 800 nm (Fig. 1).

The GAMBONS map displays the SQM and TESS-W sky brightness in magnitudes per square arcsecond computed from the radiance of each sky pixel in the respective passband, evaluated in photons/s/m2/sr and using an AB magnitude scale, whose zero points are indicated in Table 1. The zenith magnitude indicator provided by GAMBONS is calculated from the weighted average of the sky radiance over the angular field-of view of each instrument, which have rotationally symmetric Gaussian shapes with FWHM of 20° and 17°, respectively. Note that the sky magnitudes displayed by the commercial units of these devices are based on factory-set zero points chosen to mimic the magnitudes that a Johnson V detector would provide. These displayed commercial magnitudes do not have necessarily to be equal to the rigorous AB ones provided by GAMBONS.

*3.5. RGB bands*

The use of commercial digital cameras equipped with RGB sensors to develop scientific applications in astronomy has been gaining momentum in the last years (see for instance [21] or [9]). The recent definition of a standard photometric system for scientific radiometry with RGB filters and the corresponding catalogue of reference stars by [22] opens the possibility to perform accurate photometry to everyone with a commercial Digital Single Lens Reflex (DSLR) camera. Off-the shelf digital cameras have been demonstrated to be useful and reliable instruments for recording the radiance of artificial light sources both from ground [23] and space platforms like the International Space Station [24-27], as well as for quantitatively monitoring the artificial brightness of the night sky [28-30].

In this scenario, it is important to have the reference value provided by GAMBONS for the natural sky brightness in those photometric bands. The RGB passbands, radiometric quantities, and zero-points included in our model are those defined by [22].

*3.6. Human vision bands*

The scotopic $V'(\lambda)$ and photopic $V(\lambda)$ spectral sensitivity functions describe the response of the human eye fully adapted to low (< 0.005 cd m$^{-2}$) and high (> 5 cd m$^{-2}$) environmental luminances, respectively. These photometric bands have been standardized by [31-32]. The luminance of a radiance field seen with scotopic/photopic adaptation, in cd m$^{-2}$, is obtained by multiplying the integrated radiance within the $V'(\lambda)$ / $V(\lambda)$ band (in W m$^{-2}$ sr$^{-1}$) by the scotopic/photopic luminous efficacy factors K'm=1700 lm W$^{-1}$ and Km=683 lm W$^{-1}$, respectively.

To evaluate the visual brightness of the night sky it is crucial to use the adequate human sensitivity bands. As discussed by [33], the 'visible' $V$ Johnson band does not provide an accurate estimation of the visual brightness of the sky. Skies with the same brightness in $V$ Johnson may correspond to very different luminances, depending on the spectral composition of the sky radiance. The GAMBONS tool evaluates the unaided eye sky brightness natively in the human visual bands, as directly deduced from the Gaia photometry and without intermediate steps in the Johnson $V$ band, with the exception of a small number of sources in the Hipparcos catalogue.

In GAMBONS we specify the visual brightness of the night sky in conventional luminance units, cd m$^{-2}$. This brightness, however, could also be expressed in magnitudes per square arcsecond in a visual magnitude system based on either the scotopic $V'(\lambda)$ or the photopic $V(\lambda)$ passbands, depending on the adaptation state of the observers and the extent of their visual field. This possibility has already been described for photopic observers in [34]. Table 1 provides the required reference radiances in the AB scale. The magnitudes per square arcsecond thus defined could be considered truly visual ones. Recall that in case of fixating a star or sky field in foveal vision, the photometric band of choice is the photopic one, $V(\lambda)$, independently from the adaptation state of the observer, given the absence of rods and S-cone photoreceptors in the central fovea of the human eye. These photoreceptors play however a significant role in peripheral (averted) vision [34].

Work is in progress for extending GAMBONS as a tool for the ecological study of the impact of light pollution on different species, evaluating the natural sky brightness in their specific visual sensitivity bands or action spectra, accounting for their different way of perceiving and reacting to the night sky [35].

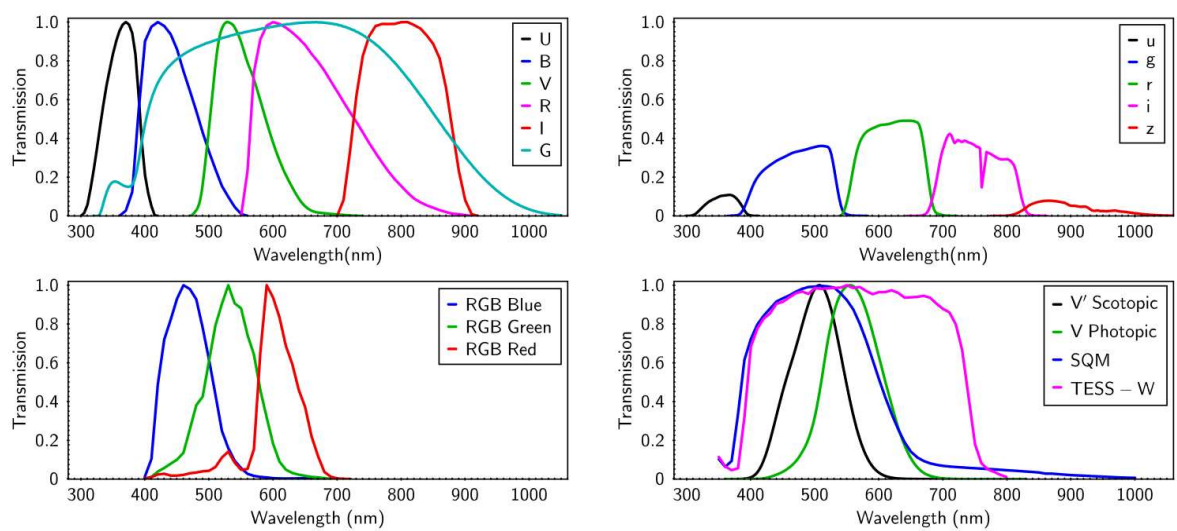

Figure 1. Passband transmissions used in this work. Upper-left panel: Johnson-Cousins [36] and Gaia G [37] bands. Upper-right panel: SDSS bands. The curves include atmospheric transmission at 1.2 air masses at the altitude of Apache Point Observatory [13]. Bottom-left panel: RGB system [22]. Bottom-right panel: V' scotopic [31], V photopic [32], SQM and TESS-W [19][18] bands.

## 4. Sky Brightness Indicators

There is not a unique parameter to assess the quality of the night sky in terms of darkness. A common indicator is the mean sky brightness of a region around the zenith. However, the zenith magnitude does not reflect the quality of the night sky in other lines of sight, that are of particular relevance in studies on the visual perception of the nightscape by humans and other species, as well as for determining the levels of artificial light affecting bands of interest for astrophysics at other zenith distances.

From the all-sky radiance maps, additional synoptic indicators of the sky brightness can be derived. In the following, we describe the ones included in this update of the GAMBONS tool.

### 4.1. Magnitude of the average radiance between two zenith angles

As a rule of thumb, the night sky is more polluted by artificial lights near horizon than at the zenith. It is therefore relevant to have indicators of the reference value of the sky brightness at different zenith angles. A useful indicator is the average in-band radiance of the sky between two zenith angles, $z_l$ and $z_u$, that is defined within the corresponding photometric band as:

$$\langle L \rangle_{(z_l, z_u)} = \frac{1}{A_{(z_l,z_u)}} \int_{\phi=0}^{2\pi} \int_{z=z_l}^{z=z_u} L(z,\phi) \sin z \, dz \, d\phi \qquad (3)$$

where $A_{(z_l,z_u)}$ is the area of the sky (in sr) between $z_l$ and $z_u$.

For a discrete sampling of the sky, as in GAMBONS, the integral can be replaced by a sum:

$$\langle L \rangle_{(z_l, z_u)} \approx \frac{1}{\sum_i \Delta_i} \sum_i L_i \, \Delta_i \qquad (4)$$

where $\Delta_i$ is equal to the HealPix (**H**ierarchical **E**qual **A**rea iso**L**atitude **Pix**elation) solid angle element (in sr) and the sum is for all the HealPix elements between $z_l$ and $z_u$.

The magnitude, in units of mag arcsec$^{-2}$, corresponding to this average radiance is computed using the usual expression:

$$m_{(z_l,z_u)} = -2.5 \log \frac{\langle L \rangle_{(z_l,z_u)}}{L_r} \qquad (5)$$

being $L_r$ the reference zero-point radiance in the given band. Note that $m_{(z_l,z_u)}$ is not the average magnitude in this region of the sky, but the magnitude of the average radiance. The non-linear relationship between radiance and magnitudes per square arc second precludes identifying both means, excepting for trivial cases.

*4.2. Global Horizontal Irradiance*

The Global Horizontal Irradiance (E) is the total amount of radiation received from above by a surface horizontal to the ground. It is measured in irradiance units (W m$^{-2}$ or photon s$^{-1}$m$^{-2}$) within the corresponding photometric band, and computed from:

$$E = \int_{\phi=0}^{2\pi} \int_{z=0}^{z=\pi/2} L(z,\phi) \cos z \sin z \, dz \, d\phi \qquad (6)$$

As in the previous case, we replace the integral by a sum:

$$E \approx \sum_i L_i \, \cos z_i \, \Delta_i \qquad (7)$$

with the sum for all the HealPix elements above the horizon.

For the human photopic and scotopic bands, *E* must be multiplied by 683 lm W$^{-1}$ or 1700 lm W$^{-1}$, respectively, to get the corresponding Global Horizontal Illuminances in lux.

*4.3. Average Upper Hemisphere Radiance*

The Average Upper Hemisphere Radiance ($L_{\text{AUHR}}$), measured in radiance units (W m$^{-2}$ sr$^{-1}$ or photon s$^{-1}$m$^{-2}$sr$^{-1}$), is the direct average of the radiance across the upper hemisphere within the corresponding photometric band:

$$L_{\text{AURH}} = \langle L \rangle_{\Omega = 2\pi} \qquad (8)$$

To get the corresponding photopic and scotopic luminances, in cd m$^{-2}$, multiply the photopic $V(\lambda)$ and scotopic $V'(\lambda)$ in-band radiances by the same factors as above.

*4.4. Average full-sphere radiance*

This is the average of the radiance calculated taking into account not only the upper hemisphere above the observer but also the light from the sky reflected by the ground (the lower hemisphere). To simplify, we assume the ground reflects light in a Lambertian way, that is, the reflected $L_R$ radiance is isotropic and is related to the incident ground horizontal irradiance $E$ (defined above) by:

$$L_R = \frac{\rho}{\pi} E \qquad (9)$$

where $\rho$ is the in-band terrain reflectance. Since the Lambertian radiance is homogeneous across the terrain and isotropic, it is equal to its average value $L_R = \langle L_R \rangle$. The average radiance of both hemispheres, $\langle L \rangle_{4\pi}$, is then:

$$\langle L \rangle_{\Omega=4\pi} = \frac{1}{2}\left(\langle L \rangle_{2\pi} + \frac{\rho}{\pi} E\right) \qquad (10)$$

All the above indicators can equivalently be specified as in-band average spectral radiance/irradiance densities, by dividing them by the passband area $\mathcal{A}$ in nm.

## 5. Web Version

A web version of GAMBONS is available at https://gambons.fqa.ub.edu. It has all the capabilities described in this paper, but it uses a simplified model for the evaluation of the effect of atmospheric scattering in order to speed up the on-line computations.

The radiance coming from a given direction of the sky has two different components: on one side, there is the direct radiance, not scattered between the source to observer; on the other hand there is the scattered radiance. The last is the radiance interacting with the atmospheric constituents located along the line of sight and scattered into the observer's field of view, adding to the detected radiance. Denoting by L0(λ, αs) the extra-atmospheric radiance from a differential patch of the sky of solid angle $d\omega$ around the generic direction αs, the total radiance scattered into the observer line of sight, Ls (λ, αs, h) can be calculated as

$$L_s(\lambda, \alpha, h) = \int_\Omega \Psi(\lambda, \alpha, \alpha_s, h)\, L_0(\lambda, \alpha_s)\, d\omega \qquad (11)$$

where $\Psi(\lambda, \alpha, \alpha_s, h)$ is the function describing the spectral radiance scattered toward the observer along the direction $\boldsymbol{\alpha}$, per unit $d\omega$, due to a unit amplitude radiant source located at αs (all directions measured in the observer reference frame). $h$ is the observer altitude above sea level. For our model, we have used the first order scattering $\Psi(\lambda, \alpha, \alpha_s, h)$ function described in Kocifaj-Kránicz [38] with an effective scattering phase function composed of aerosol and molecular (Rayleigh) components weighted by the corresponding optical depths. The integral in Eq. 11 is a double integral of a function extended to the whole hemisphere above the observer $\Omega$. Even for low spatial resolution maps (i.e. one square degree per pixel), it is a computationally expensive operation, not feasible to implement in a response-immediate web service.

Whereas we use Eq. 11 for routine sky brightness calculations, we have implemented a simpler approach in the web version of our model in order to alleviate the computational load. It consists in replacing the aerosol (A) and molecular (M) optical depth $\tau_0^{A/M}(\lambda)$ by an effective optical depth $\tau_{0,eff}^{A/M}(\lambda) = \gamma\, \tau_{0,}^{A/M}(\lambda)$ with $\gamma < 1$. The value of $\gamma$ depends on the aerosol albedo and asymmetry parameter [39]. This is the approach used for diffuse sources in [40]. The effective optical depth cannot exactly reproduce the scattering model described by Eq. 11 using a single $\gamma$ value, but it is a practical option for obtaining reasonably accurate results when computing time is a constraint.

As an example, we show in Fig. 2 the differences between the results obtained from both models when using $\gamma = 0.5$, $\tau_0^A = 0.15$ and three different bands (B, V and R Johnson-Cousins). As a general trend, the simplified model overestimates the brightness of the natural sky near the horizon in all the bands, while it underestimates the brightness at zenith. At bright areas of the Milky Way both models give very similar results. The results of the comparison depends on the band, as well as on the other parameters like the aerosol and molecular optical depth or the intensity of the airglow. It is expected that the simplified model differ less than 0.1 magnitude per $arcsec^{-2}$ for the most of the cases, with the trend shown in Fig. 2.

If more accurate results are required, the complete scattering model using Eq. 11 is implemented in the stand-alone version of GAMBONS, available under request to the authors.

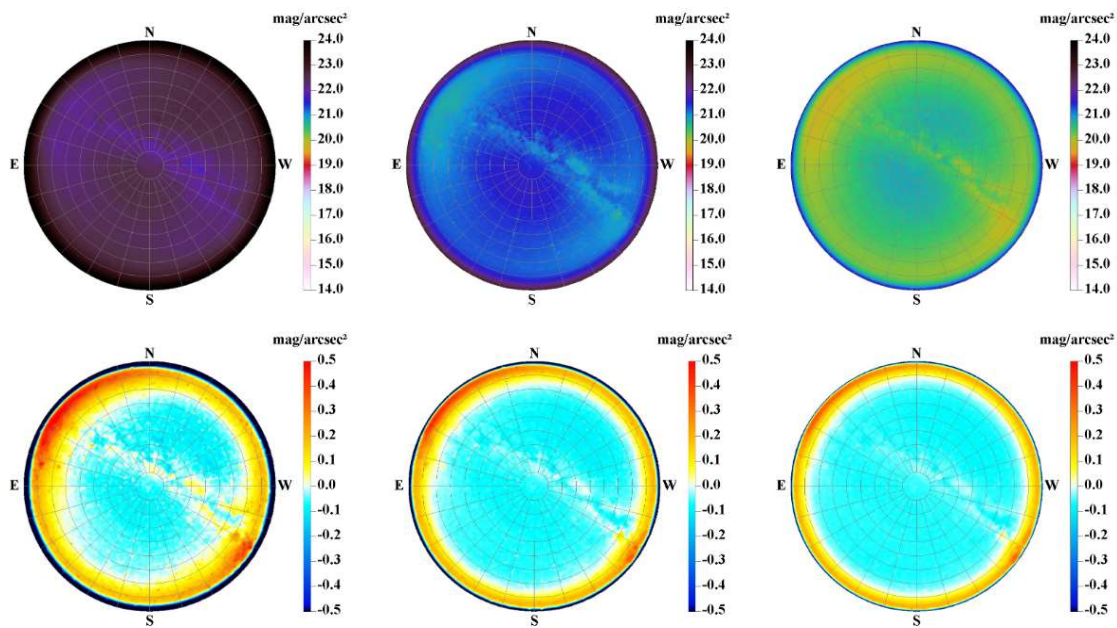

Figure 2. Comparison of the scattering model described by Eq. 11 with the simplified scattering model (γ = 0.5) used in the web version of GAMBONS. Top panels: natural sky brightness map in B, V and R Johnson-Cousins bands. Bottom panels: difference map (Eq. 11 minus simplified model) for each band. Units are mag $arcsec^{-2}$.

## 6. GAMBONS applications

GAMBONS is intended to be a tool to get the reference value of the natural night sky brightness in cloudless and moonless conditions in any given photometric band, if reliable transformations from Gaia and Hipparcos photometry can be obtained. The first product of the model are all-sky maps that allows to easily visualize the sky brightness. A sample of these maps are shown in the top panels of Fig. 2.

But the model has other direct applications. For instance, it also allows to compare the relative weight of the different components contributing to the natural sky brightness. In particular, we have considered the integrated star light, the background light, including both the diffuse galactic light and the extragalactic background light, the zodiacal light and the airglow. Paper I describes in detail these components and how are they computed by the model. Table 2 summarizes the relative contribution of different components in the five Johnson-Cousins bands for a region of the sky of radius 10° around the zenith, for a latitude ϕ = 40° N at sea level. The zenith brightness was computed at one hour intervals from the end to the beginning of astronomical twilight, and then averaged for one year to obtain the tabulated values. The aerosol optical depth $\tau_0^A$ was set to 0.15, with an Ångstrom exponent equal to 1. The values for the three parameters of the scattering function $\Psi(\lambda, \alpha, \alpha_s, h)$ in Eq. 11 were chosen to be the following: the aerosol $\omega_A$ and molecular $\omega_M$ albedos were set to 0.85 and 1, respectively; and the asymmetry parameter $g$ to 0.9. The model was run with a reference airglow spectrum computed from ESO's SkyCalc web interface, calculated for the Cerro Paranal altitude (2640 m above the sea level) in the wavelength range from 350 nm to 1050 nm, and with a value of the Monthly Averaged Solar Radio Flux equal to 100 sfu (1

sfu = $10^{-22}$ W m$^{-2}$ s$^{-1}$), the approximate average value in the period 2009-2020 (one solar cycle) according to the data of the Canadian Space Weather Forecast Centre (CSWFC). Any change in these parameters will modify, to a greater or lesser extent, the results in Table 2, that must to be considered only strictly valid for this particular set of parameters, which could be modified by the GAMBONS user, if required. It is noticeable the great differences that exist among different bands. Of all the components, airglow is showing the largest absolute dependence with wavelength. While in the $U$ and $B$ band there are almost no emission lines and the airglow radiance corresponds to the continuum, the number and strength of the lines increases from 558 nm (OI green line, almost at the maximum of the $V$ band transmission) and it is particularly important in the $R$ and $I$ bands. On the other hand, the spectrum of the zodiacal light is basically the solar spectrum (slightly reddened), and the spectrum of the integrated star light shows its peak in the $R$ band, with faint emission in the $U$ band. Altogether modules the relevance of the different contributors to the sky brightness in the different bands. This band dependence shows the need of multi-band studies when evaluating the effects of the light pollution.

Other application of GAMBONS is to study the annual variation of the natural night sky brightness in a given place due to the different area of the celestial sphere visible along the year. Taking into account this intrinsic variability is crucial when assessing the quality of the night sky from field measurements. The presence or absence of the Milky Way, or the height of the zodiacal light over the horizon must be carefully considered when interpreting the measurements. In Fig. 3 we show the mean zenith natural sky brightness at local midnight during a year for the five Johnson-Cousins bands at $\phi$=40° N, for the particular case of the parameters described above.

The variability of the atmospheric conditions and the airglow intensity, at almost at any time scale from minutes to years, is an essential factor determining the natural night sky brightness and must be taken into account as an intrinsic dimension of this problem. The very concept of the existence of a 'natural' or 'reference' night sky brightness value characteristic of any given observation site, much less if intended to be valid for the whole Earth, has no operative meaning unless the particular state of the atmosphere, or the statistical mix of states used to estimate it, are explicitly described in the definition of this indicator. The possibility to set several atmospheric parameters, as the aerosol and molecular optical depth, the Ångstrom exponent and the airglow intensity, allows GAMBONS to reproduce the sky brightness under a wide set of atmospheric conditions.

The radiance outside the Earth's atmosphere, including the integrated starlight, the background light and the zodiacal light, is available online. This data supersedes the data published in Paper I.

Table 2. Percentage (%) of the mean contributions to zenith natural night sky brightness for a latitude equal to 40° N for Johnson-Cousins bands.

| Component | $U$ | $B$ | $V$ | $R$ | $I$ |
|---|---|---|---|---|---|
| Airglow | 64.6 | 35.7 | 44.0 | 55.9 | 77.0 |
| Integrated Star Light | 20.8 | 34.3 | 27.2 | 21.7 | 12.4 |
| Zodiacal Light | 12.4 | 26.3 | 25.3 | 19.6 | 9.0 |
| Background Light | 2.2 | 3.7 | 3.5 | 2.8 | 1.6 |

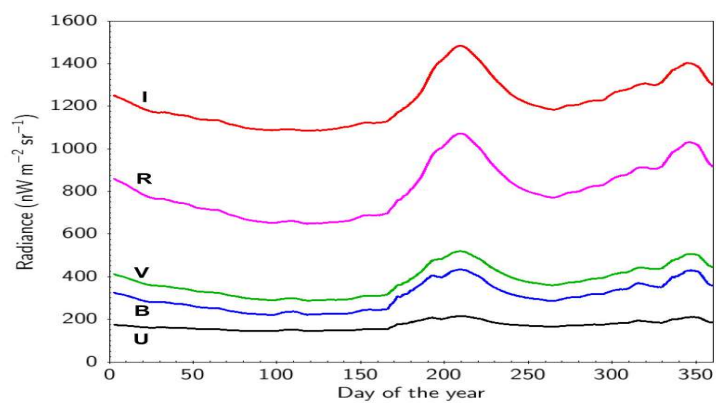

Figure 3. Midnight radiance at zenith in the Johnson-Cousins bands as function of the day of the year for an observer at latitude equal to 40° N.

## 7. Conclusions

We have described an updated version of the Gaia map of the brightness of the natural sky (GAMBONS). The main novelties are the use of the Gaia EDR3 data release and the inclusion of new photometric bands. Some of these bands, as the ones in the RGB system, are expected to be of growing importance in the coming years in the light pollution studies, thanks to the standardization work by [22]. Furthermore we have added new sky brightness indicators, as the global horizontal irradiance, the average upper hemisphere radiance and the average full-sphere radiance. These new indicators help to assess the quality of the night sky in terms of darkness, beyond the usual mean sky brightness around the zenith.

The simplified scattering model used in the GAMBONS web version has been compared with the more detailed model described in paper I. The expected differences between both models are less than 0.1 mag arcsec$^{-2}$, being the simplified model darker at zenith and brighter near the horizon. However, this result depends on the considered band, the aerosol and molecular optical depth and of the value of $\gamma$ parameter. The results show that the simplified model is a reasonable option when fast computation is needed and no highly accurate results are required. A stand-alone version of GAMBONS, with the complete scattering model is available under request to the authors.

Finally, we have shown some applications of the model. Particular examples of the relative contributions to the natural night sky brightness of the different components and the annual variation of the sky brightness for a given place were computed for the five bands of the Johnson-Cousins photometric system. It is shown the different behaviour of the different bands, revealing the need of focusing the study of the light pollution in a multi-band approach.

**Acknowledgements**

This work was supported by the MINECO (Spanish Ministry of Economy) through grant RTI2018-095076-B-C21 (MINECO/FEDER, UE). Authors at the ICC-UB acknowledge financial support from the State Agency for Research of the Spanish Ministry of Science and Innovation through the "Unit of Excellence María de Maeztu 2020-2023" award to the Institute of Cosmos Sciences (CEX2019-000918-M). SB acknowledges Xunta de Galicia, grant ED431B 2020/29.